\documentclass[square]{ws-procs11x85}
\usepackage{balance} 

\newcommand{\met}{\hbox{E\kern-0.5em\lower-0.1ex\hbox{/}}_T}

\begin{document}

\twocolumn[
\title{A three stage model for the inner engine of GRBs: Prompt emission and
early afterglow}

\author{J. Staff}

\address{Purdue University, Department of Physics,
525 Northwestern Avenue,
West Lafayette, IN 47907-2036, USA}

\author{B. Niebergal and R. Ouyed}

\address{University of Calgary,
SB 605,
2500 University Drive NW,
Calgary, Alberta, Canada,
T2N 1N4}

\begin{abstract}
We describe a model within the `Quark-nova'' scenario to interpret the
recent observations of early X-ray afterglows of long Gamma-Ray Bursts (GRB)
with the Swift satellite. This is a three-stage model within the context of
a core-collapse supernova. STAGE 1 is an accreting (proto-) neutron star
leading to a possible delay between the core collapse and the GRB. STAGE 2
is accretion onto a quark-star, launching an ultrarelativistic jet
generating the prompt GRB. This jet also creates the afterglow as the jet
interacts with the surrounding medium creating an external shock. Slower
shells ejected from the quark star (during accretion), can re-energize the
external shock leading to a flatter segment in the X-ray afterglow. STAGE 3,
which occurs only if the quark-star collapses to form a black-hole, consists
of an accreting black-hole. The jet launched in this accretion process
interacts with the preceding quark star jet, and could generate the flaring
activity frequently seen in early X-ray afterglows. Alternatively, a STAGE
2b can occur in our model if the quark star does not collapse to a black
hole. The quark star in this case can then spin down due to magnetic
braking, and the spin down energy may lead to flattening in the X-ray
afterglow as well. This model seems to account for both the energies and the
timescales of GRBs, in addition to the newly discovered early X-ray
afterglow features.
\end{abstract}
\vskip12pt  
]

\bodymatter

\section{Introduction}

\begin{figure}
\includegraphics[width=0.49\textwidth]{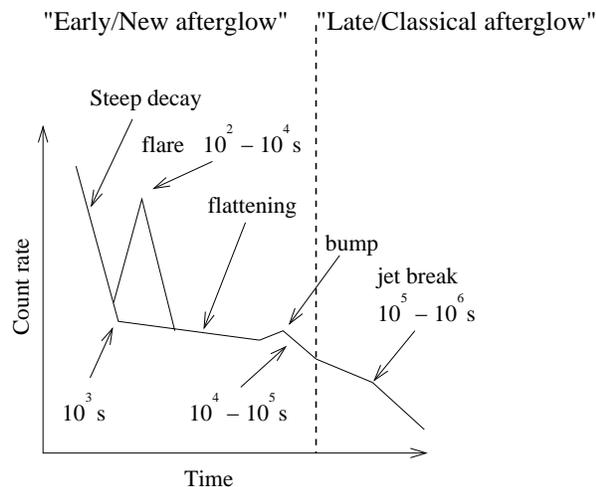}
\caption{Generic X-ray afterglow (e.g. \cite{zhang06}). A steep decay often
follows the prompt emission. Then a flat segment is commonly seen, before
the afterglow decays following one or two power laws. The flat segment can
also be followed by a steep decay (not shown), indicating that the flat
segment was caused by inner engine activity. Such flat segment with steep
decay is the topic of this paper. Overlaid on the initial steep
decay and the flat segment is sometimes one or more flares. }
\label{genericag}
\end{figure}

Prior to the launch of Swift, the gamma ray burst (GRB) afterglow was thought 
to follow a power-law (or possibly
two power laws, where the break is due to the jet break\cite{rhoads97}).
However, Swift showed that the X-ray afterglow light curve starting from
about 100 seconds after the GRB trigger in many cases cannot be described by
a power law. Instead, it can be divided into several phases, as shown in
Fig.~\ref{genericag}. Not all stages are present in all bursts. The first
stage is a very sharp drop off, thought to be due to the curvature
effect\cite{kumarpanaitescu00}. Thereafter a flatter segment
lasting for $10^4$ to $10^5$ seconds is often seen\cite{obrien06}, followed 
by the two power
laws mentioned earlier (denoted ``Late/Classical afterglow'' in Fig.~\ref{genericag}).
In some cases this flat segment is followed by a very sharp drop
off, before the classical afterglow starts. Overlaid 
on the initial sharp drop off and on the flat segment can be
one or more flares.

The flaring activity and flat segments followed by
a sharp drop off in the X-ray afterglow is a very strong indication
that the engine is still powering the light curve\cite{troja07}. Therefore the inner
engine must not only be able to explain the energetics and duration of the
 prompt gamma ray emission, but also long lasting features in the afterglow. 

In this paper we discuss how a quark star (QS) as the engine can explain these 
observed features.
In our model, the prompt emission is caused by internal shocks\cite{piran99} in an
ultrarelativistic jet, launched by accretion onto a QS. Flares are
also caused by internal shocks, either by continued accretion onto the
QS, or if the QS collapsed to a black hole (BH), by accretion
onto a BH.
A flat segment not followed by a sharp drop off can be explained by refreshed 
shocks\cite{reesmeszaros98}, i.e. slower shells emitted during the
prompt phase slowly catching up with the external shock refreshing it, or by
extended engine activity. A sudden cessation in the extended engine activity
can explain a flat segment followed by a sharp drop off in the observed
light curve. Extraction of the QS spin energy
can explain such a long lasting engine activity. If the QS collapses to a BH
during spin down, the engine activity will end suddenly and a sharp drop off
can be seen.

We will describe the model in section~\ref{threestagessection},
and in particular discuss how a flat segment followed by a sharp drop off
sets an upper limit on the magnetic field, if one assumes that the flat
segment is due to extraction of spin energy. On the other hand, the quark
star magnetic field must be strong enough to channel the accreting material
to the polar cap, something that puts a lower bound on the magnetic field.
In section~\ref{070110section} we will take GRB 070110 as an example (a
burst that show a flat segment followed by a sharp drop off), and use the
theory from section~\ref{threestagessection} to show that we can
consistently account for both the energetics and duration of the prompt
phase and the plateau phase. We summarize in section~\ref{summarysection}.

\section{Three stages model}
\label{threestagessection}

\begin{figure}
\includegraphics[width=0.49\textwidth]{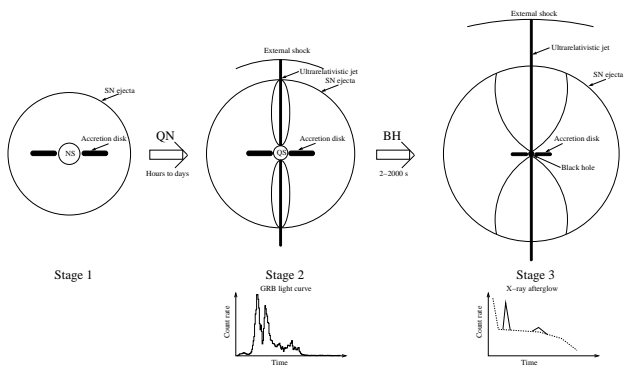}
\caption{The three stages model for a GRB inner engine. {\it Stage 1:} A  
neutron star is formed in the core collapse of a massive star.
{\it Stage 2:}
An ultrarelativistic jet launched from hyperaccretion onto a QS 
generates the GRB.
{\it Stage 3:} 
An ultrarelativistic jet from a BH collides with slower parts of the
QS jet and creates flaring.
}
\label{threestagesfig}
\end{figure}

\subsection{Stage 1}

In \cite{staff07}, a three stage model for long GRBs was
proposed (see Fig.~\ref{threestagesfig}).  Stage
1 is a (proto-) neutron star phase, the neutron star being born in the
collapse of the iron core in an initially massive star. This neutron star
can be transformed to a QS, either through spin-down \cite{staff06} or through
accretion, thereby increasing its central density sufficiently that
it can form strange quark matter. This stage could lead to
a delay between the core collapse and the GRB.  The collapse into a QS, in a
quark nova\cite{ouyed02} (QN), releases up to $10^{53}$ ergs that might
help power the explosion of the star. This can possibly explain why GRB
associated supernovae are often very energetic\cite{ouyed07,
leahy07}, and depending on the size of the stellar envelope this might also
create a GRB precursor\cite{ouyed07}.
 If a QS is formed directly in the core collapse,
stage 1 will be bypassed and the process starts from stage 2. The same
amount of energy will still be emitted from the conversion to a QS,
allowing these features even if stage 1 is bypassed.

\subsection{Stage 2}

Stage 2 is the QS stage. 
Accretion onto the QS surrounded by a hyperaccreting debris disk
from the collapse launches a highly variable
ultra-relativistic jet, in which internal shocks can give rise to the gamma
radiation seen in a GRB \cite{ouyed05}. In order for a jet to form, the
accreting material has to be channeled to the polar cap. This sets a lower
limit on the QS magnetic field, in that the Alfv{\'e}n radius has to
be at least twice the radius of the star. The Alfv{\'e}n radius is given
by:
\begin{equation}
r_{\rm A}=\bigg(\frac{B^2 R_{\rm QS}^6}{\dot{m}\sqrt{2GM_{\rm QS}}}\bigg)^{2/7}.
\label{msradius}
\end{equation}
In this equation G is the gravitational constant, $M_{\rm QS}$ is the QS mass,
$R_{\rm QS}$ is the QS radius, $\dot{m}$ is the accretion rate which is 
constrained by observed energetics of the burst, and B is the QS magnetic field.
Later we will find the maximum magnetic field in order to explain the flat
segment in GRB 070110, and we can use this together with the Alfv{\'e}n
radius to find an estimate for $\dot{m}$, and compare this to what is
required to explain the prompt emission.

If the QS survives the accretion phase and is rapidly rotating, a stage 2b
can be reached in which 
magnetic braking spins the star down extracting about $10^{52}$ ergs
in rotational energy. Magnetospheric currents escapes through
the light cylinder, thereby launching a secondary outflow and spinning down
the QS, whose magnetic field is aligned with the rotation axis. 
The spin down luminosity\cite{niebergal06} is given by:
\begin{equation}\label{eq:qssd_lum}
L \sim 4\times 10^{48} ~{\rm erg~s}^{-1} \left(\frac{B_0}{10^{15}~{\rm
G}}\right)^2 \left(\frac{2~{\rm ms}}{P_0}\right)^4 \left(1 + \frac{t}{\tau}
\right)^{-5/3} \ ,
\end{equation}
where the characteristic spin-down time (in seconds) is,
\begin{equation}
\tau = 3.5\times 10^3~{\rm s} \left(\frac{10^{15} {\rm G}}{B_0}\right)^2 
                    \left(\frac{P_0}{2 {\rm ms}}\right)^2 
                    \left(\frac{M_{\rm QS}}{1.4M_{\odot}}\right)
                    \left(\frac{10 {\rm km}}{R_{\rm QS}}\right)^4 \ .
\label{sdtime}
\end{equation}
In the above equations t is time since the beginning of spin down, $P_{0}$ is the initial
spin period, and $B_{0}$ is the initial magnetic field strength.  

There will be a break in the spin down luminosity at $t=\tau$. After the
break the spin-down luminosity will follow a power law with power $-5/3$,
however, this might not be the observed light curve power. If, on the other
hand, this secondary outflow created by spinning down the QS ends
abruptly, a much sharper drop off will be seen, as the light curve drops
down to the level given by the external shock. This can occur if the quark
star collapses to a BH during spin down. Unless a disk exists around
the BH, the rotational energy of the BH cannot be extracted
and the engine has been shut off. 

\subsection{Stage 3}

If the
QS (in the process of ejecting the shells) accreted a sufficient amount of
matter, it can collapse into a BH, which leads to stage 3.
Stage 3 is accretion onto the BH which launches
another ultra-relativistic jet \cite{devilliers05}.  Interaction
between this jet and the 
jet from the QS can give rise to flaring commonly seen in the X-ray
afterglow of GRBs. Internal shocks within the BH jet itself can also
lead to flaring activity. Alternatively, if the QS did not collapse to a BH,
continued accretion onto the QS after the prompt phase might also explain
X-ray flaring. The BH jet might be very powerful, so if it
catches up with the external shock a bump might be seen in the light curve.
 The relevant features and emission have been discussed in details in
\cite{staff07}.

\section{Application of our model to GRB 070110}
\label{070110section}

\begin{figure}
\includegraphics[width=0.35\textwidth,angle=270]{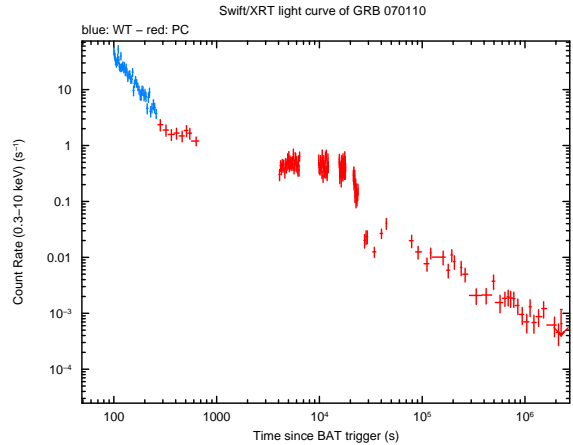}
\caption{The X-ray afterglow light curve of GRB 070110\cite{evans07}. A flat 
segment is
clearly seen lasting up to about 20000 seconds (about 6000 seconds when
corrected for cosmological time dilation), followed by a very sharp decay.
The flat segment we explain as being due to extraction of spin energy from a
rapidly rotating QS. The QS collapsed to a BH,
ending the secondary outflow created by the spin down and leading to the sharp
drop off in the light curve, as the light curve drops down to the level
given by the external shock.}
\label{070110xrtfig}
\end{figure}

Based on the observed duration of the flat segment in a burst, we use
Eq.~\ref{sdtime} to find the magnetic field required to spin the star down
in this time (assuming that the star collapsed to a BH at $t=\tau$).
This will give an upper limit to the magnetic field, since if the magnetic
field is weaker, $\tau$ is larger than the time of collapse. Having an
estimate for the magnetic field, we can go on to estimate the spin down
luminosity using Eq.~\ref{eq:qssd_lum}. This spin down luminosity can be
compared to the observed luminosity. We assume that $10\%$ of the spin down
luminosity can be converted into X-ray photons. Furthermore, in order to
explain the prompt emission, we can estimate the accretion rate that this
magnetic field can channel to the polar cap. Since the estimate of the
magnetic field gave a maximum, this leads to a maximum possible accretion
rate. We now apply this method to GRB 070110, a burst that show a very
sharp drop off following a flat segment (see Fig.~\ref{070110xrtfig}).

The maximum magnetic field found by setting $t=\tau$ is $6.8\times10^{14}$ G.
The corresponding maximum spin down luminosity
is $1.6\times10^{48} {\rm erg/s}$. The observed luminosity during the flat
segment is $L_{\rm Obs, iso}\approx10^{48} {\rm erg/s}$\cite{troja07}. If we assume a 10
degree opening angle for this outflow, this corresponds to $L_{\rm Obs,
10}\approx1.5\times10^{46} {\rm erg/s}$. Even when we assume that $10\%$ of
the spin down luminosity goes into radiation, we see that we have 10 times
more energy than required to explain this burst. 

A magnetic field of $6.8\times10^{14}$ G can channel
$7.5\times10^{-4}M_\odot/{\rm s}$ (Eq.~\ref{msradius}) to the polar cap. 
If we assume that $1\%$
of the total gravitational energy of the accreted material is converted to
prompt radiation, this corresponds to $3.8\times10^{50}$ erg in prompt gamma
radiation. Assuming that the beaming angle of this GRB was $10$ degrees, the
observed gamma ray energy in this burst was $E_\gamma\sim1\times10^{50}$
erg. Again, we get more energy than required in order to explain this burst.
The reason for this can be that we have overestimated the magnetic field,
i.e. the QS collapsed at $t_{\rm collapse}<\tau$, that we have
overestimated the efficiency with which gamma radiation is produced, that the
majority of the energy is radiated at frequencies not observed by SWIFT, 
or that the beaming angle is larger.

\section{Summary}
\label{summarysection}

The properties of the three stages (initiated by the collapse of the iron
core in an initially massive star) of the inner engine in our model are:
\begin{itemize}
\item If a neutron star is left behind, it can collapse to a quark-star
at a later stage, creating a delay between the supernova and the GRB.
\item Accretion onto the quark-star generates the
GRB by powering an ultrarelativistic jet. Internal shocks in this jet
create the GRB.
\item Accretion continues, but at some point it cannot heat up the star
sufficiently. This halts the emission of shells, ending the GRB.
\item If the quark-star accretes a sufficient amount of matter, it collapses
into a black-hole. Further accretion onto the black-hole launches an
ultrarelativistic jet.
\item If the QS does not collapse to a black-hole and is rapidly
rotating, magnetic braking of the QS can extract the rotational
energy. This process can last for $10^4-10^5$ seconds. About $10^{52}$ ergs
can be extracted this way.
\end{itemize}
The emission features in our model can be summarized as follows:
\begin{itemize}
\item Early, steep decay in the X-ray afterglow is due to the curvature
effect.
\item Flares are created by interaction between the jet from the
accretion onto a black-hole and slower parts from the jet from the
quark-star.
\item Re-energization of the external shock (seen as flattening of the
X-ray afterglow light curve) is due only to the jet from the quark-star.
Slower parts of this jet re-energizes the external shock.
\item If spin-down energy from the QS is extracted, this outflow
(mainly in the form of $e^+e^-$ pairs) can scatter photons from the
external shock creating a flat segment in the X-ray afterglow, see
\cite{panaitescu07}.
\item If a black-hole is formed the jet from the black-hole can collide with
the external shock from the GRB, creating a bump in the afterglow light
curve.
\end{itemize}

In conclusion, we have presented a three stage model for the inner engine
for GRBs. This model seems to be able to account for the observed prompt
gamma ray emission, as well as the features of the early X-ray afterglow.

\balance

\end{document}